# Improving Access to Digitized Historical Newspapers with Text Mining, Coordinated Models, and Formative User Interface Design

**Robert B. Allen**
College of Information Science and Technology
Drexel University
Philadelphia, PA 19107 USA
+1-215-895-0460
rba@drexel.edu

**Abstract:**
Most tools for accessing digitized historical newspapers emphasize relatively simple search; but, as increasing numbers of digitized historical newspapers and other historical resources become available, we can consider much richer modes of interaction with these collections. For instance, users might use exploratory search for looking at larger issues and events such as elections and campaigns or to get a sense of "the texture of the city… how the city was thinking." To take full advantage of rich interface tools, the content of the newspapers needs to be described systematically and accurately. Moreover, collections of multiple newspapers need to be richly cross-indexed across titles and even with historical resources beyond the newspapers.

## 1. Introduction

Because an increasing number of digitized full-text historical newspapers is now available, we can begin to shift from searching them one title at a time to considering the way access to several titles can be coordinated. For instance, in the time frame 1880 to 1920 for Washington DC about ten different titles are available. Similarly, at the state level newspapers from several cities are being digitized and there should be synergies among them.

Collections such as the LC/NEH National Digital Newspaper Program (NDNP) collection are particularly useful for detailed text processing because they provide public domain OCR along with word coordinates and font identification in the META-ALTO format. However, because of factors such as the poor quality of the originals and the necessity of reproduction from microfilm copies, the OCR is of uneven quality. Moreover, there are additional aspects of the newspapers that are not coded in the ALTO files.

Allen et al. (2008) explored a "pipeline" processing model with a series of stages for creating article-level metadata. One step in this processing is segmentation of the page image. This segmentation is based on the identification of large fonts in the text which indicates a headline and the top of an article. The text of each of the segments was then categorized by genre and topic categories based on the standard developed by the International Press and Telecommunications Council (http://www.iptc.org). The metadata assignment was moderately successful for narrow categories which included highly distinctive terms but it was less successful for categories which depended on nuanced language processing. Allen and Hall (submitted) explored regular news features which were not captured by the original set of genre categories.

In summary, it is easy to process some categories which are associated with distinctive keywords automatically but many other categories are less accurately processed even though most of the content is highly predictable based on factors such as its location in an issue. The more expert knowledge added the higher the accuracy. However, the amount of digitized newspapers is so large that it does not seem feasible for people, even groups of citizens engaged in collaborative correction, to make all the corrections needed. It seems unlikely that complete corrections can be accomplished by the automated process but automated processing can augment the capabilities of the human beings.

## 2. Text Mining and Modeling History

Text mining can be used to identify patterns in the text which should also be useful for improving the text processing. For instance, Allen et al. (2008) reported finding a seasonal pattern for the word "drought". In fact, such regularities are easy to find and here we present rich data for two additional examples. These are drawn from the *Philadelphia Evening Ledger* from mid-September to December 31, 1914. As shown in Figure 1, occurrences of the term "Thanksgiving" peak at Thanksgiving and mentions of the term "Christmas" peak, unsurprisingly, at Christmas. Moreover, other terms related to the holidays such as "turkey" and "Santa" follow similar patterns.

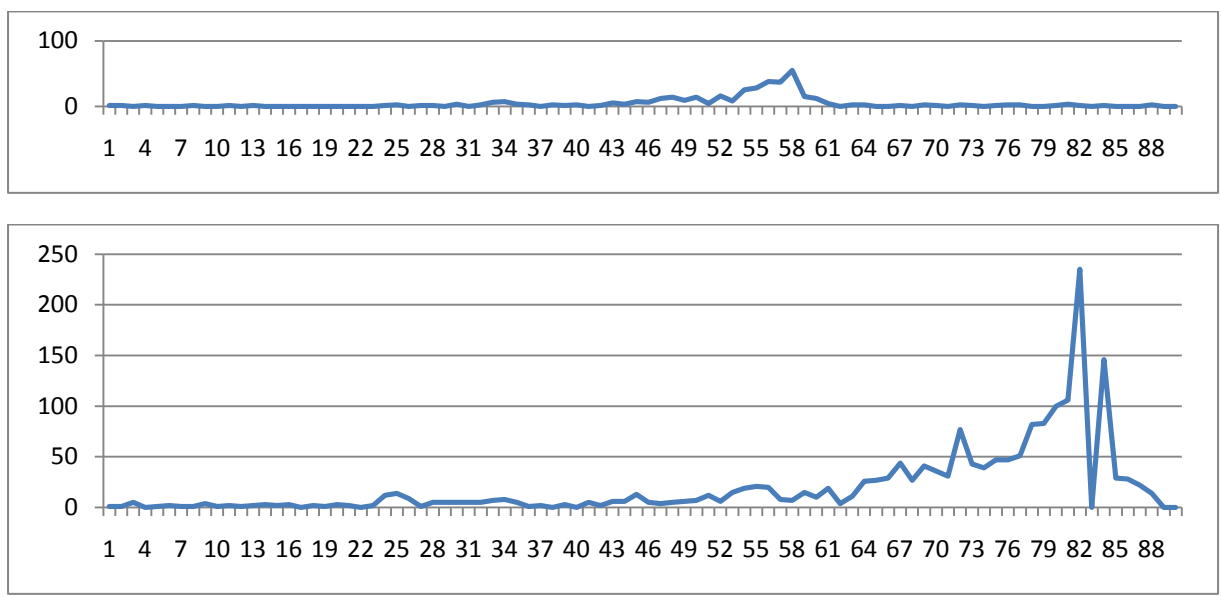

**Figure 1: Frequencies for the term "Thanksgiving" (upper panel) and "Christmas" (lower panel).**

Clearly, there are patterns in the data and they could be useful in making corrections in the text. For instance, they could be used to set weights for OCR corrections. However, to truly understand and exploit them it will be more helpful to develop models which account for and could even predict them. Thus, we want to move from simply observing such regularities to modeling them. Models could be based on many factors. For instance, we would predict very different types of news reports from a rural community than from an urban one. That is, we might develop what we could call "community models" (Allen et al, 2007). These would provide a unified framework for interrelating the people, places, organizations, and events that appear in the newspaper for a town or city.

Local reporting has a context within national and international reporting, even though readers at the time may not have direct access to non-local views. From a historical perspective there is, then, a requirement to enable the correlation of local news to the wider context and indeed to tracking the spread of news in times when local news readers did not have ready access to alternative news sources, such as telephones, television, radio or the Internet. A focus on local context also facilitates access to data that enables understanding of local social networks, social activities, sports, entertainment, community government functioning, advertising norms, and biographic data. These local contexts can then be compared across geography and time to analyze local and regional dispersion of news. In addition, to enhance understanding of the significance of search results, particularly from multiple newspapers, data such as local census data, economic data, or weather data, would be made available for overlaying on the search results. Moreover, relevant named-entities can be derived from many other sources. The "community models" that we propose incorporate these functions. Ultimately, these also need to be combined with newspaper models which encapsulate the editorial, stylistic, and production policies of each newspaper.

In this case, we are particularly interested in determining accessing civic processes, by which we mean government-related activities, which, of course comprise a substantial portion of the news. Figure 2 once again show data from the *Philadelphia Evening Ledger* for the fall of 1914. In this case, the data emphasizes terms related to the election of 1914 which was held on the first Tuesday of November. This is of particular interest because it shows a progression from the campaign to the election. Thus, we have identified a strong sequence of events by examination of word frequencies. This pattern is confirmed with terms such as "candidate" and "rally" for the days leading up to the election and the terms "election", "votes", and "voted" for the days surrounding the election. Obviously, these are very gross measures but they do clearly demonstrate the predictable evolution of events and, thus, we may think of them as being a model of events that generate news.

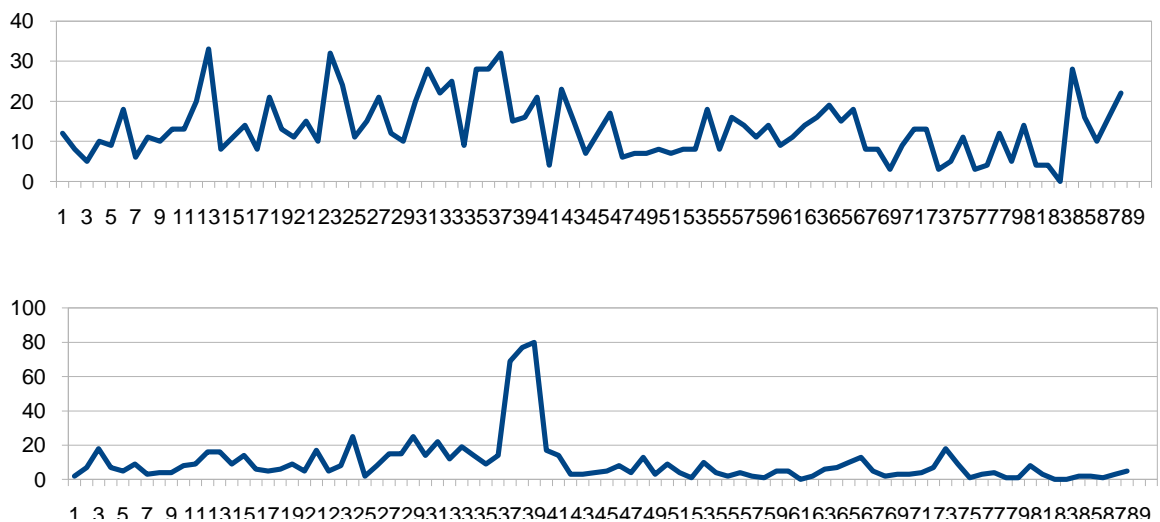

**Figure 2: Frequency of the term "campaign" (upper panel) and of the word "vote" (lower panel). Note that term "campaign" is highest in the days leading up the election while term "vote" has a peak only for the few days surrounding the election.**

## 3. Preliminary Results for Text Mining Concurrent Local Newspapers

While we have a fairly complete record for significant national events, we have a much less complete record for important events in individual communities. Such a record would be of interest in its own right in addition to being potentially useful for adding constraints for the text processing. However, identifying significant events from individual newspapers can be difficult because of the large amount of material to sift through. One strategy for focusing on significant events would be to compare coverage from different newspapers. There are a few cases where there are multiple newspapers digitized. In particular, the Library of Congress itself has processed a number of historical newspapers for the District of Columbia (DC). Of particular interest, there is a period of about three months in 1906 for which we have digitized samples of two major Washington DC newspapers: the *Washington Times* and the *Washington Herald*.

We collected the OCR output from the first pages of the newspapers during the period of the overlap. We cleaned that OCR by identifying only those words which were also found among the words in the Associated Press and *New York Times* portions of the Linguistic Data Consortium Gigaword corpus[1]. That is, we tried to minimize the OCR errors by comparing the OCR text to words from a very large sample of English text. In addition, a stop list of the 500 most common words in English was also applied. Using the document-frequency measure from information retrieval, we identified terms which showed a striking change in frequency from their overall baseline. Finally, we found those words which showed that change of frequency in both Washington DC newspapers within a three-day window. These words are often indicators of distinctive news stories. Table 1 shows three examples selected from the output of this process.

| Oct 29 1906 | awful breaking bridge camden coach dempsey drawbridge heroism motorman picked submerged surface survivors thoroughfare trestle windows |
|---|---|
| Nov18 1906 | colon dillon hopes lacking princeton princetons teams tigers yale |
| Dec 31 1906 | ambulances awful belt coaches cotta crowded empty horribly identified mangled relief rescuers splintered takoma terra |

**Table 1: Examples of distinctive terms for news stories which appeared across two different newspapers in Washington DC on about the same day in late 1906. This technique allows us to identify news stories of particular significance because they appear in both newspapers.**

These are preliminary results and much more work needs to be done to make them robust and useful. For instance, the mention of "colon" in the second line refers to the city of Colon, Panama and is unrelated to a main story detected about the defeat of Yale by Princeton in an American football game.

### 4. Toward a Historian's Workbench

While dedicated researchers can and do exhaustively examine the rich resources of newspapers with microfilm, modern user interfaces for digitized materials should make the job easier for everyone and the barriers to entry much lower for beginning researchers. Thus, it is time to consider how historians might interact with a much richer set of materials than they have previously been able to do. That is, we might think of developing a historian's workbench (cf., Toms & Flora, 2006). Robert Sieczkiewicz, the Drexel University Archivist, and I are conducting interviews with historians to determine the features historians would find particularly useful (Allen et al. 2010).

One set of issues concerns searching itself. For instance, one historian said. "The *[existing commercial online]* database is good for searching names, but broader topics are hard to research". Another researcher said she used newspapers to fill in gaps in research and corroborate information from other sources. Her exploratory searching included looking at larger issues and events such as elections and campaigns. That is, rather than searching on specific items, she used newspapers to find public opinion about issues such as changes in liquor license laws – to get a sense of "the texture of the city… how the city was thinking". That is difficult to do with simple keyword indexes of the materials.

---

[1] http://www.ldc.upenn.edu/LDC2003T05, Jan 28, 2003.

Another set of issues deals with managing results from searches. One of the historians interviewed said "a log of all searches – this is a huge issue for me". When editing a book manuscript recently, she found it "hugely taxing" to find items she hadn't cited. Similarly, "searches lead to other searches", so she would like ways to see how searches are nested within each other and to return to earlier search results. She also asked for "a visual map telling you where you are in your search" as well a system that lets her easily use multiple windows.

Similar comments are also found in blogs on the Web. "Rachel"[2], who describes herself as a doctoral student in history, presents a set of techniques for searching newspapers.
> (1) Have a List,
> (2) Find a thread of some kind,
> (3) Don't just use one newspaper,
> (4) Don't fall into the trap of only reading articles that your keywords throw up.
> (5) Use existing secondary literature,
> (6) Keep really, really scrupulous notes, and
> (7) Don't neglect the letters and the advertisements.

Taken together with the interviews, these may suggest that a user interface should have both flexible searching and also tools for annotation and management of the search results.

## 5. Conclusions

We have presented the value of extending text processing and text mining with modeling of underlying activities that are reported in a newspaper. Indeed, we can view newspapers as a type of projection of, or perhaps a filter for, the community they are reporting about. In some cases these models can be based on predictable patterns. But, in other cases they are so difficult to predict that it is best to try to identity events as they arise. Thus, we also report results from comparing the contents of two newspapers from the same city for the same time frame to find stories that are reported in both of them.

While we have explored many attributes of the historical newspapers, they have so much more rich material that we have barely begun to get a complete picture. Moreover, the application of findings such as these will be a huge undertaking. But, the result will be ready access to these exceptionally rich historical resources. Indeed, we envision digitized newspapers cross referenced with a wide range of other historical resources including primary sources as letters and secondary sources such as textbooks. Ideally, they would be woven into an automatically generated historical narrative.

## 6. Bibliography and References

---

[2] Rachel, A Historian's Craft: http://idlethink.wordpress.com/2009/06/16/on-newspapers-as-sources/ (accessed November 2009)